\documentclass[a4paper,10pt,twoside]{cpc-hepnp}
\usepackage{multicol}
\usepackage{graphicx}
\usepackage{booktabs}
\usepackage{amssymb,bm,mathrsfs,bbm,amscd}
\usepackage[tbtags]{amsmath}
\usepackage{lastpage}

\begin{document}

\fancyhead[c]{Submitted to `Chinese Physics C'}

\title{Compact narrow-band THz radiation source based on photocathode rf gun\thanks{Supported by National Natural Science Foundation of China (11205152) }}

\author{%
      LI Wei-Wei
\quad HE Zhi-Gang$^{1)}$\email{hezhg@ustc.edu.cn}%
\quad JIA Qi-Ka} \maketitle

\address{%
National Synchrotron Radiation Laboratory, University of Science and
Technology of China, Hefei, 230029, Anhui, China
}

\begin{abstract}
Narrow-band THz coherent Cherenkov radiation can be driven by a subpicosecond electron bunch traveling along the axis of a hollow cylindrical dielectric-lined waveguide. We present a scheme of compact THz radiation source based on the photocathode rf gun. On the basis of our analytic result, the subpicosecond electron bunch with high charge (800pC) can be generated directly in the photocathode rf gun. A narrow emission spectrum peaked at 0.24 THz with 2 megawatt (MW) peak power is expected to gain in the proposed scheme (the length of the facility is about 1.2 m), according to the analytical and simulated results.
\end{abstract}



\begin{multicols}{2}

\section{Introduction}
Terahertz (THz) radiation, which lies in the frequency gap
between the infrared and microwaves, typically referred to as the
frequencies from 100 GHz to 30 THz, is finding use in an increasingly wide
variety of applications\cite{lab1}. Electron bunches of subpicosecond duration can be
utilized to generate high power, coherent, THz radiation. Recent examples of powerful THz radiation include generation via synchrotron radiation (SR)\cite{lab2,lab3,lab4}, transition radiation (TR)\cite{lab5}, Smith-Purcell radiation (SPR)\cite{lab6}, and Cherenkov radiation (CR)\cite{lab7}. As another candidate of THz radiation source, THz-FEL is also being developed worldwide\cite{lab8,lab9,lab10}, including the table-top FEL devices\cite{lab11,lab12}.

The coherent Cherenkov radiation (CCR)\cite{lab13} can be excited by the passage of relativistic electron bunch through a hollow cylindrical dielectric tube coated on the outer surface with metal, which is to form a dielectric-lined waveguide (DLW). It is normally used for dielectric wakefield accelerator\cite{lab14} and Cherenkov free electron laser (CFEL)\cite{lab15}. The CCR wakefields are confined to a discrete set of modes due to the waveguide boundaries. This slow-wave structure supports modes with phase velocity equal to the electron beam velocity that are thus capable of efficient energy exchange with the beam. For a given driving electron bunch, the dimensions of DLW structure should be carefully chosen such that the narrow-band THz CCR can be excited in the structure\cite{lab7} and microbunches with a periodicity of THz wavelength can be formed with the use of a chicane\cite{lab16}.

The photocathode rf gun is normally used to provide electron beam with high brightness, which is given by~$B_n=I/(\varepsilon_{n,x}\cdot \varepsilon_{n,y})$~. The~$I$~is the peak current, ~$\varepsilon_{n,x}$~and~$\varepsilon_{n,y}$~are the normalized transverse emittances in two directions, respectively. Because the peak current can be increased by bunch compression with a chicane and the transverse emittance is preserved irreversibly in the accelerator, so the transverse emittance is the most important parameter in the traditional applications. If the transverse emittance is not required strictly, the electron bunch with higher peak current can be obtained in the gun by carefully choosing the operation parameters of the drive laser spot and the gun, detailed demonstrations can be found in our previous work\cite{lab17}. Therefore, high peak power narrow-band THz radiation can be excited in a compact facility based on the photocathode rf gun (the longitudinal size is about 1.2 m). By using a train of femtosecond drive laser pulses with THz frequency, a compact THz-FEL scheme is also proposed\cite{lab12,lab18}, whose size is equivalent compared to our scheme.

We introduce a brief summary of the theory of wakefields in a cylindrical dielectric-lined waveguide in Section 2. In Section 3, a scheme of compact narrow-band THz radiation source is proposed which is based on the photocathode rf gun. In this section, we discuss the generation of subpicosecond electron bunch with relative high charge, analyse the impact of the form factors of electron bunch on the CCR wakefields, and present the analytical and simulated results of the THz radiation. Section 4 presents a summary of this paper.

\section{Theory of wakefields in a cylindrical dielectric-lined waveguide}
The structure of DLW is shown as Fig. 1.
\begin{center}
\includegraphics[width=5.0cm]{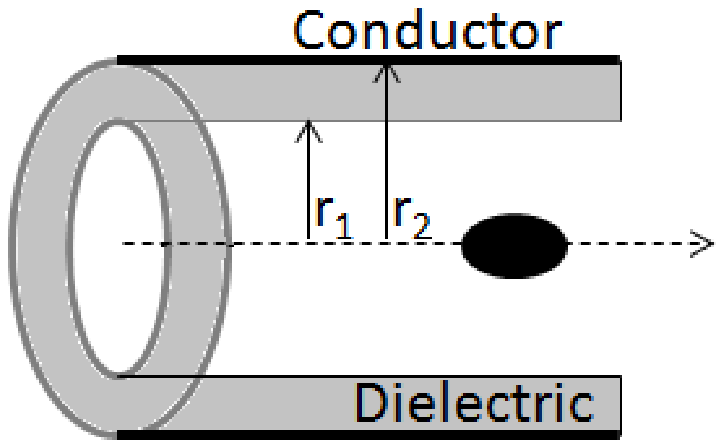}
\figcaption{\label{f1}The structure of DLW.}
\end{center}
Fourier expansion of the longitudinal fields in a circular cylindrical waveguide takes the form:
\begin{equation}
\begin{array}{c}
 \left( \begin{array}{l}
 {E_z}(r,t) \\
 {H_z}(r,t) \\
 \end{array} \right) = \frac{1}{{{{(2\pi )}^3}}}\int_{ - \infty }^\infty  {d\omega dk}  \times  \\
 \sum\limits_{l =  - \infty }^\infty  {\exp [ - i(\omega t - kz - l\theta )]}  \times \left( \begin{array}{l}
 {e_z}(r) \\
  - i{h_z}(r) \\
 \end{array} \right) \\
 \end{array}
\end{equation}
For a given~$l$~in Eq. (1), the eigenmodes can be obtained with the boundary conditions. The eigenfunctions will be designated by subscripts such as~${e_{z,n}}(r)$~,~${h_{z,n}}(r)$~and the eigenvalues by~${\omega _n}$~,~${k_n}$~, etc.
For the~$T{M_{0n}}$~modes, the dispersion equation describing the transverse modes of the DLW structure is given by\cite{lab19}
\begin{equation}
\frac{{{I_1}({k_{0n}}{r_1})}}{{{I_0}({k_{0n}}{r_1})}}=\frac{{{\varepsilon _r}{k_{0n}}}}{{{\kappa _{on}}}}\frac{{{J_0}({\kappa _{on}}{r_2}){Y_1}({\kappa _{on}}{r_1}) - {Y_0}({\kappa _{on}}{r_2}){J_1}({\kappa _{on}}{r_1})}}{{{Y_0}({\kappa _{on}}{r_1}){J_0}({\kappa _{on}}{r_2}) - {Y_0}({\kappa _{on}}{r_2}){J_0}({\kappa _{on}}{r_1})}}
\end{equation}
where~${k_{0n}} = \sqrt {{k_n}^2 - {{\left( {{\omega _n}/c} \right)}^2}}$~is the radial wave numbers in the vacuum region and~${{\kappa _{0n}} = \sqrt {{\varepsilon _r}{\mu _r}{{\left( {\frac{{{\omega _n}}}{c}} \right)}^2} - {k_n}} }$~is the radial wave numbers in the dielectric region. ~${{\varepsilon _r}}$~ is the relative permittivity of the material, and~$n = 1,2,3 \ldots$~indexes the solutions to the transcendental equation. ~${J_m}(x)$~ and~${Y_m}(x)$~are Bessel functions of the first and second kinds of order m, and~${I_m}(x)$~is the modified Bessel function of the first kind. For a given driving electron bunch with the velocity~$\beta c$~(~$\beta c = \frac{{{\omega _n}}}{{{k_n}}}$~), the dimensions of DLW tube~$({r_1},{r_2})$~can be chosen such that only the~$T{M_{01}}$~mode is coherently excited, thus selecting a single operating frequency.

For the azimuthally symmetric transverse distribution of electron bunch, only~$T{M_{0n}}$~modes are excited. In our case, the influence of the transverse distribution of electron bunch on the electromagnetic field is weak (the proof will be presented in Section 3.3 ), so only the temporal distribution of electron bunch is considered. The orthonormality relation between any two eigenmodes and radiative power flow can be written as\cite{lab20}
\begin{equation}
\sum\limits_{i = 1}^{N = 2} {\int_{{R_{i - 1}}}^{{R_i}} {dr \cdot r} } \left[ {{\varepsilon _i}{e_{z,m}}\left( r \right){e_{z,n}}\left( r \right) + {\mu _i}{h_{z,m}}\left( r \right){h_{z,n}}\left( r \right)} \right]={C_n}{\delta _{mn}}
\end{equation}
\begin{equation}
\overline {{P_{0n}}}=-\beta cq_0^2 \frac{{e_{z,n}^2\left( 0 \right)}}{{{C_n}}}\Theta(- s)\cdot{\alpha _n}^2
\end{equation}
where~$q_0$~is the charge,~$\Theta ( - s)$~means the radiation is excited behind the electron. The~${\alpha _n}$~is the form factor and defined by
\begin{equation}
{\alpha _n} = \left| {\int\limits_{ - \infty }^\infty  {dsf\left( s \right){e^{ - i{k_n}s}}} } \right|
\end{equation}
Where~$f\left( s \right)$~is the longitudinal distribution function of the electron bunch and~$\int {dsf\left( s \right)}  = 1$~.

The pulse length and energy of the THz radiation are given by
\begin{equation}
{t_{pulse}} = \frac{L}{{{v_g}}} - \frac{L}{{\beta c}}
\end{equation}
\begin{equation}
U=\frac{L}{{\beta c}}\cdot \overline {{P_{0n}}}
\end{equation}
where the~$v_g$~is the group velocity of the radiation,~$L$~is the length of DLW structure. The magnitude of the electric field on the longitudinal axis is given approximately by\cite{lab20}
\begin{equation}
{\left[ {{E_z}(r = 0)} \right]_{0n,\max }} \cong  - 2{q_0}\frac{{{e_{z,n}}^2\left( 0 \right)}}{{{C_n}}}{\alpha _n}
\end{equation}
The group velocity~$v_g$~of the wakefields, representing the speed of energy flow along
the waveguide, is simply the constant of proportionality between the total power
flow and the field energy per unit length.Each waveguide mode has its own group velocity,and for the~$T{M_{0n}}$~mode
\begin{equation}
{v_g} = \frac{{\int_A {{S_{z,0n}}dA{}_z} }}{{\int_A {\beta {U_{0n}}dA{}_z} }}
\end{equation}
where ~$U = \left( {1/8\pi } \right)\left( {\varepsilon {\bf{E}} \cdot {\bf{E}} + \mu {\bf{H}} \cdot {\bf{H}}} \right)$~is the electromagnetic energy density, and ~${S_z} = \frac{c}{{4\pi }}\left( {{E_r}{H_\theta } - {E_\theta }{H_r}} \right)$~is the Poynting vector.

\section{Scheme of compact narrow-band terahertz radiation source}
\begin{center}
    \includegraphics[width=8.0cm]{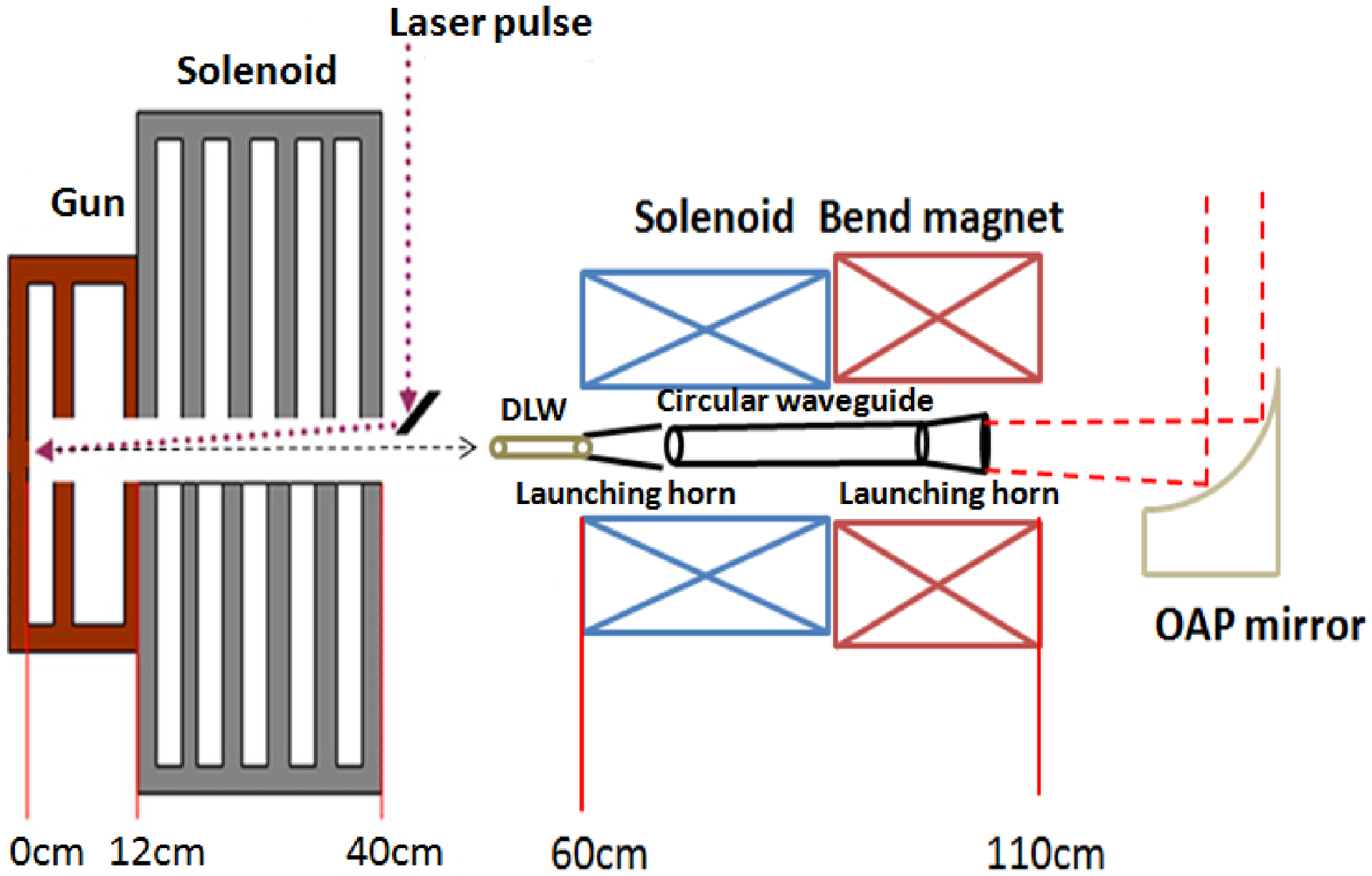}
    \figcaption{\label{f2}Layout of the proposed THz radiation facility.}
    \end{center}
Fig. 2 is the layout of the proposed THz radiation facility. The subpicosecond electron bunch is generated in the photocathode rf gun and focused by the solenoid. The narrow-band THz radiation is driven by the electron bunch when it travelling through the DLW tube. The launching horn and circular waveguide is used to transport the radiation. The electron bunch is restricted by another solenoid during the transportation. Finally, a bend magnet is used to bend the electron bunch to the Faraday cup.

\subsection{Generation of the subpicosecond electron bunch in the photocathode rf gun}
The length of electron bunch is affected by such factors as space charge effect, beam energy and energy spread, and the coupling effect between the transverse and longitudinal emittances.

The bunch lengthening due to the space charge effect can be estimated in a drift space by\cite{lab21}:
\begin{equation}
\Delta\sigma_z=2qcl^2/I_ar\sigma_z\gamma^4
\end{equation}
where q is the bunch charge, c is the speed of light, l is the drift distance, ~$I_a=1.7~kA$~, r is the
bunch radius, ~$\sigma_z$~ is the bunch length and ~$\gamma$~ is the beam energy. In the photocathode rf gun, the energy of electron beam is low, so the space charge effect plays the dominant role. In order to decrease the bunch lengthening caused by the space charge effect, the acceleration gradient should be as high as possible and the radius of the drive laser spot should be chosen appropriately. To decrease the bunch lengthening caused by the coupling effect between the transverse and longitudinal emittances, laser shaping\cite{lab22} should be considered to restrain the growth of the transverse emittance. Furthermore, the bunch length can be compressed in the gun by tuning the acceleration phase\cite{lab23}.

The acceleration gradient of the BNL type gun for the LCLS is designed to operate at 140 MV/m with 120 Hz repetition rate\cite{lab24}. We consider to improve the cathode seal technique\cite{lab25} of our gun machined by the Department of Engineering Physics of Tsinghua University, and look forward to improve the acceleration gradient from the current 80 MV/m to 120 MV/m. It is already achieved at the latest generation gun in the Tsinghua University\cite{lab26}. For the 80 MV/m gradient, we can tune the bunch charge and laser spot size to obtain sub-picosecond electron bunch\cite{lab17}.  Although the uniform laser spot can be achieved by using a spatial shaper, it is difficult to transport the shaped spot to the cathode. So we plan to clip an expanded gaussian laser spot by an aperture to restrain the impact of nonlinear space charge force on the transverse emittance. To obtain a short electron bunch with relative high charge (800 pC), the radius of aperture is chosen at 4 mm. For our laser pulse (the measured rms length is about 2.0 ps), the simulated (by using code ASTRA\cite{lab27}) evolutions of the rms beam size and bunch length along the longitudinal position are shown in Fig. 3, when the acceleration phase is set at 4 degrees. Fig. 4 shows the transverse distribution and current distribution of electrons at the focal point of the first solenoid, the rms length of electron bunch is about 0.65 ps.
\begin{center}
    \includegraphics[width=7.0cm]{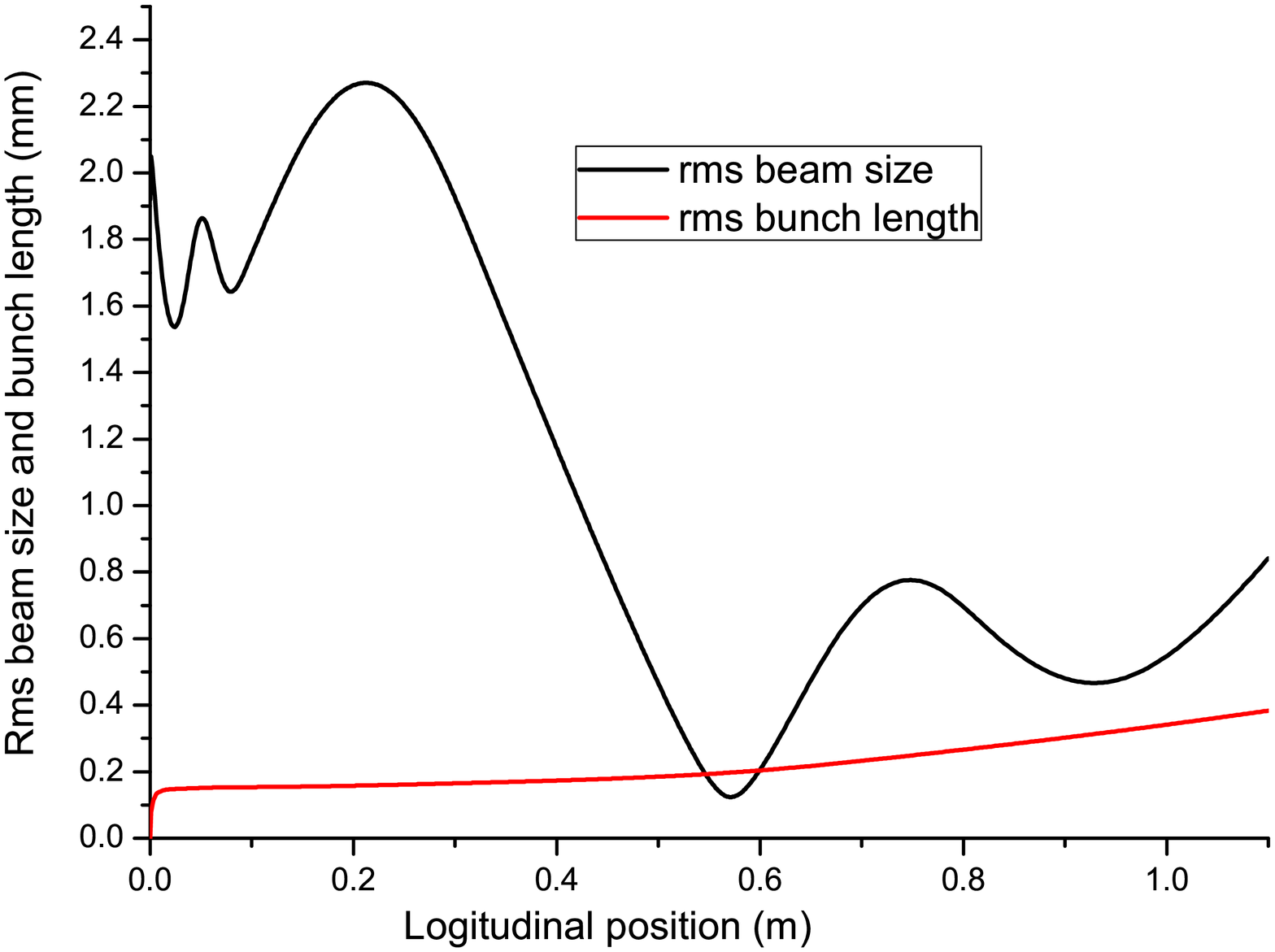}
   \figcaption{\label{f3}The rms beam size and bunch length evolutions along the longitudinal position.}
    \end{center}
\begin{center}
    \includegraphics[width=8.0cm]{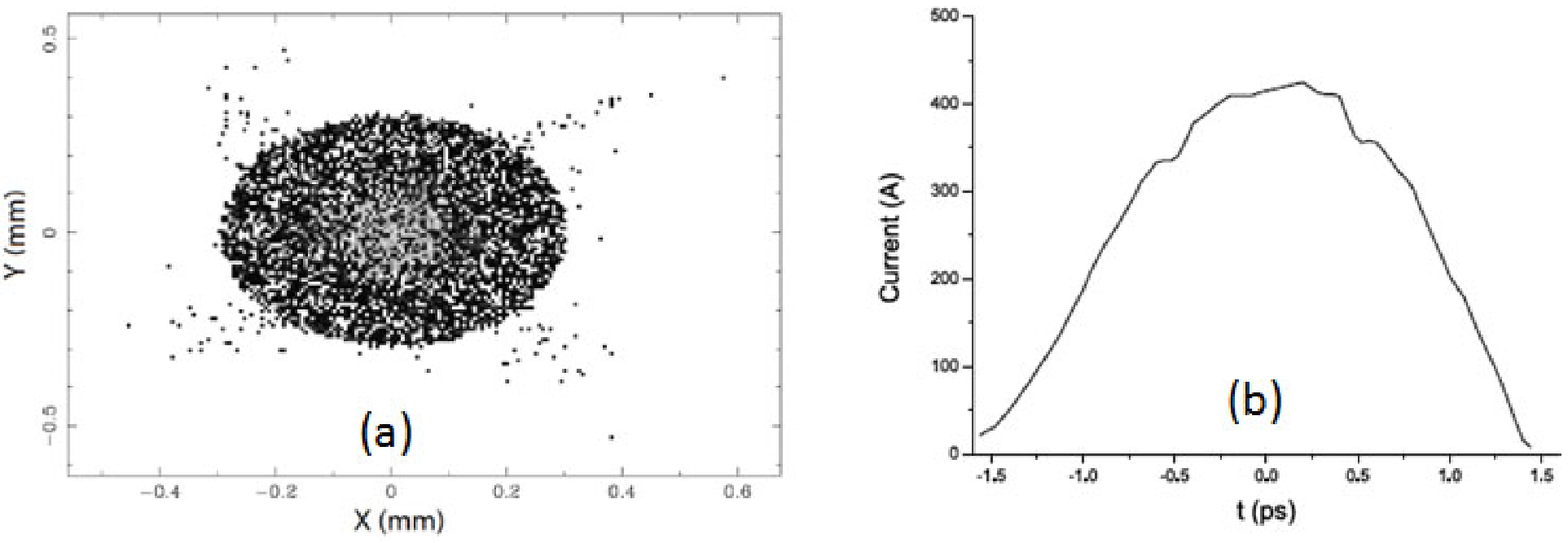}
    \figcaption{\label{f4}The transverse distribution and current distribution of electrons.}
   \end{center}

\subsection{The dimensions of DLW structure}
The material of the dielectric is fused silica, and the dielectric constant is~$\varepsilon_r=3.8$~. According to the transverse distribution of electrons shown in Fig. 4 (a), the inner radius of the DLW is chosen at~$400~\mu m$~. Based on the Eq. (2) and the electron bunch length~${\sigma _z} = 0.65~ps$~shown in Fig. 4 (b), the~$r_2$~is chosen at ~$550~\mu m$~. Then, only the~$T{M_{01}}$~mode is coherently excited, and the frequency is about~$0.241~THz$~. The~$1~cm$~length of the structure is a preliminarily choice.

\subsection{Impact of the form factors of electron bunch on the THz radiation}
In the respect of the impact of transverse form factor on the THz radiation, we start with the solution equation of the longitudinal electromagnetic field. The longitudinal components of the fields satisfy Bessel¡¯s equation
\begin{equation}
\left( \begin{array}{c}
 {{\tilde J}_z}(r) \\
 0 \\
 \end{array} \right) = \left[ {\frac{{{d^2}}}{{d{r^2}}} + \frac{1}{r}\frac{d}{{dr}} + \frac{{\varepsilon \mu {\omega ^2}}}{{{c^2}}} - \frac{{{l^2}}}{{{r^2}}}} \right]\left( \begin{array}{l}
 {e_z}(r) \\
 {h_z}(r) \\
 \end{array} \right)
\end{equation}
where
\begin{equation}
{\tilde J_z}(r) = 4\pi \int\limits_{ - \infty }^\infty  {dt} \int\limits_{ - \infty }^\infty  {dz\int\limits_0^{2\pi } {dz{e^{i(\omega t - kz - l\theta )}}} } \left[ {\frac{1}{\varepsilon }\frac{{\partial \rho }}{{\partial z}} + \frac{\mu }{{{c^2}}}\frac{{\partial {J_z}}}{{\partial t}}} \right]
\end{equation}
The fields~${e_z}(r)$~and~${h_z}(r)$~can be expanded in terms of their eigenmodes, as
\begin{equation}
\begin{array}{c}
\left( {\begin{array}{*{20}{c}}
   {{e_z}\left( r \right)}  \\
   {{h_z}\left( r \right)}  \\
\end{array}} \right) = \sum\limits_{n = 1}^\infty  {{A_n}} \left( {\begin{array}{*{20}{c}}
   {{e_{z,n}}\left( r \right)}  \\
   {{h_{z,n}}\left( r \right)}  \\
\end{array}} \right)
\end{array}
\end{equation}
Because the electron bunch is azimuthally symmetric (~$l=0$~) and only exist in the vacuum area (~$r < {r_1}$~),  the orthonormality relation Eq. (3) can be used to find the amplitudes~$A_n$~
\begin{equation}
\begin{array}{c}
{A_n} = \frac{1}{{{C_n}\left( {{k^2} - {k_n}^2} \right)\left( {{\beta ^2} - 1} \right)}}\int_0^{{r_1}} {drr{e_{z,n}}\left( r \right)} {\tilde J_z}\left( r \right)
\end{array}
\end{equation}
For the~$T{M_{01}}$~mode, the~${e_{z1}}(r)$~is a monotone increasing function of~$r$~, so if~$\Delta =\frac{{{e_{z,1}}\left( {{r_1}} \right) - {e_{z,1}}\left( 0 \right)}}{{{e_{z,1}}\left( 0 \right)}} \ll 1$~, we get the approximation~${e_{z,1}}(r < {r_1}) \approx {e_{z,1}}(r = 0)$~. In our case, the~$\Delta=0.0021$~and the value of~$\int_0^{{r_1}} {drr} {\tilde J_z}\left( r \right)$~is only relevant to the charge, therefore we can conclude that the electromagnetic fields and radiation power can be regarded as approximately independent of the transverse distribution of the electron bunch.

For a electron bunch with temporal symmetrical distribution, when the bunch length satisfies~$k_n\sigma_z\ll1$~, the temporal form factor Eq. (5) can be rewritten as
\begin{equation}
\begin{array}{c}
 \begin{array}{c}
 \alpha_n  \approx \int\limits_{ - \infty }^\infty  {dsf\left( s \right)} [1 - i{k_n}s - \frac{{{k_n}^2{s^2}}}{{2!}} + i\frac{{{{\left( {{k_n}s} \right)}^3}}}{{3!}}] \\
  = 1 - \frac{{{k_n}^2}}{2}\int\limits_{ - \infty }^\infty  {dsf\left( s \right)} {s^2} = 1 - \frac{{{k_n}^2}}{2}{\sigma _z}^2 \\
 \end{array}
 \end{array}
 \end{equation}
therefore the form factor is independent of the temporal distribution state.
In our case, the temporal form factor of electron bunch (~$\sigma_z=0.65~ps$~) shown in Fig. 4 (b) calculated by Eq. (5) is 0.604. For a gaussian distribution with~$\sigma_z=0.65~ps$~, the form factor calculated by Eq. (5) is 0.614. The difference between this two values is~$1.67\%$~.
\subsection{Simulation of the THz radiation}
We simulate the radiation by using the code xoopic\cite{lab28}, and the gaussian electron bunch is used on account of the demonstration in the Section 3.3. The parameters used in the simulation are shown in Table 1.
\begin{center}
\tabcaption{ \label{tab1}  Parameters used in the simulation.}
\footnotesize
\begin{tabular*}{80mm}{c@{\extracolsep{\fill}}ccc}
\hline
Bunch charge & Q & 800pC \\
Bunch energy  & E & 5.53~$MeV$~ \\
rms energy spread &  & 5\% \\
rms bunch length (Gaussian) & ~$\sigma_z$~ & 0.65~$ps$~ \\
rms bunch size (Gaussian) & ~$\sigma_x$~ & 110~$\mu m$~ \\
rms normalized emittance & ~$\varepsilon_n$~ & 7.5~$mm\cdot mrad$~ \\
\\
Dielectric inner radius & ~$r_1$~ & 400~$\mu m$~ \\
Dielectric outer radius & ~$r_2$~ & 550~$\mu m$~ \\
Length of the dielectric & L & 1 cm \\
Dielectric constant & ~$\varepsilon_r$~ & 3.8 \\
\bottomrule
\end{tabular*}
\end{center}
The longitudinal electric field is shown in Fig. 5,
\begin{center}
\includegraphics[width=7.0cm]{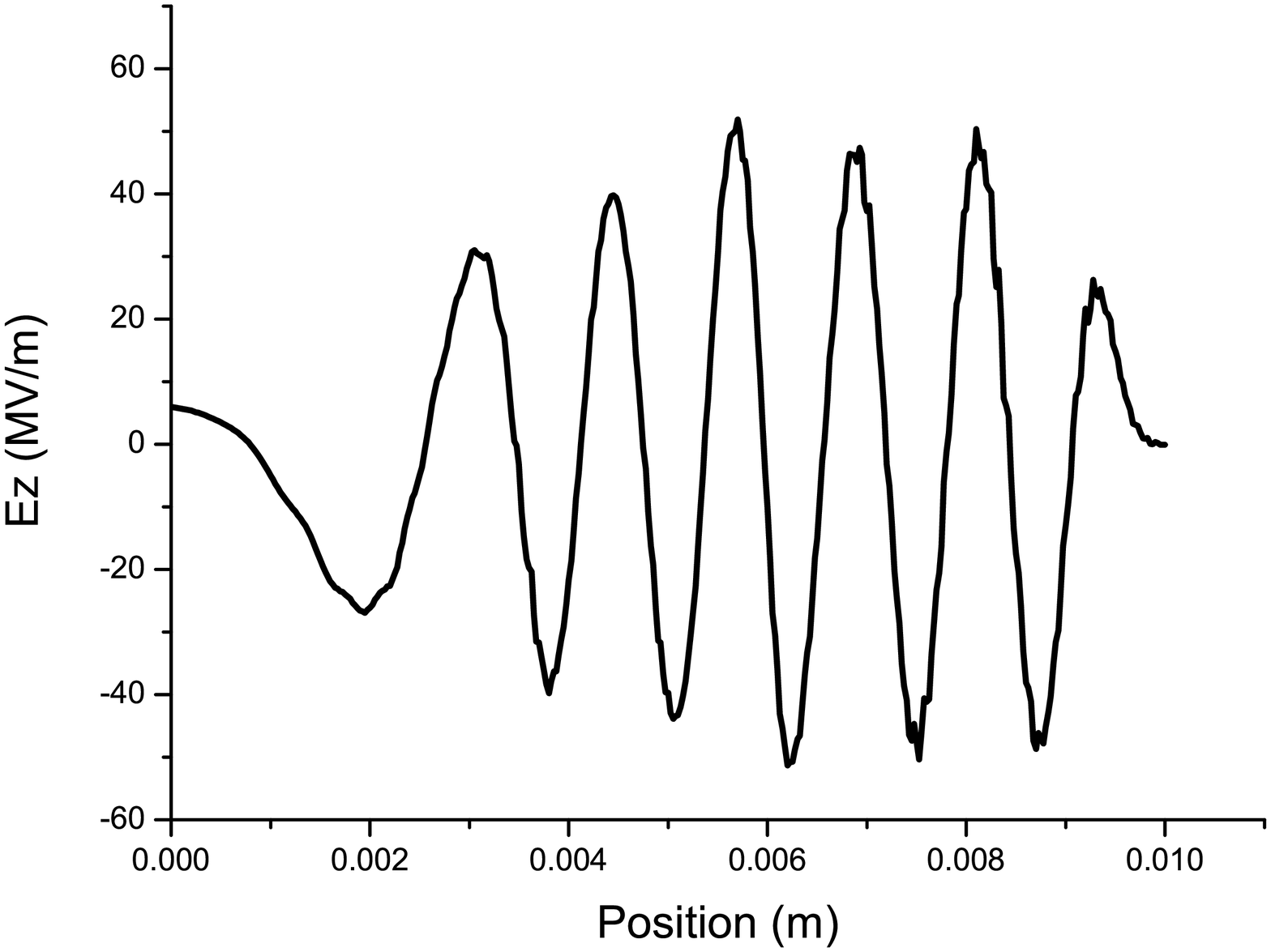}
\figcaption{\label{f5}The longitudinal electric field on the axis.}
\end{center}
and the power spectrum is shown as Fig. 6, which is calculated numerically from the longitudinal on-axis electric field.
\begin{center}
\includegraphics[width=7.0cm]{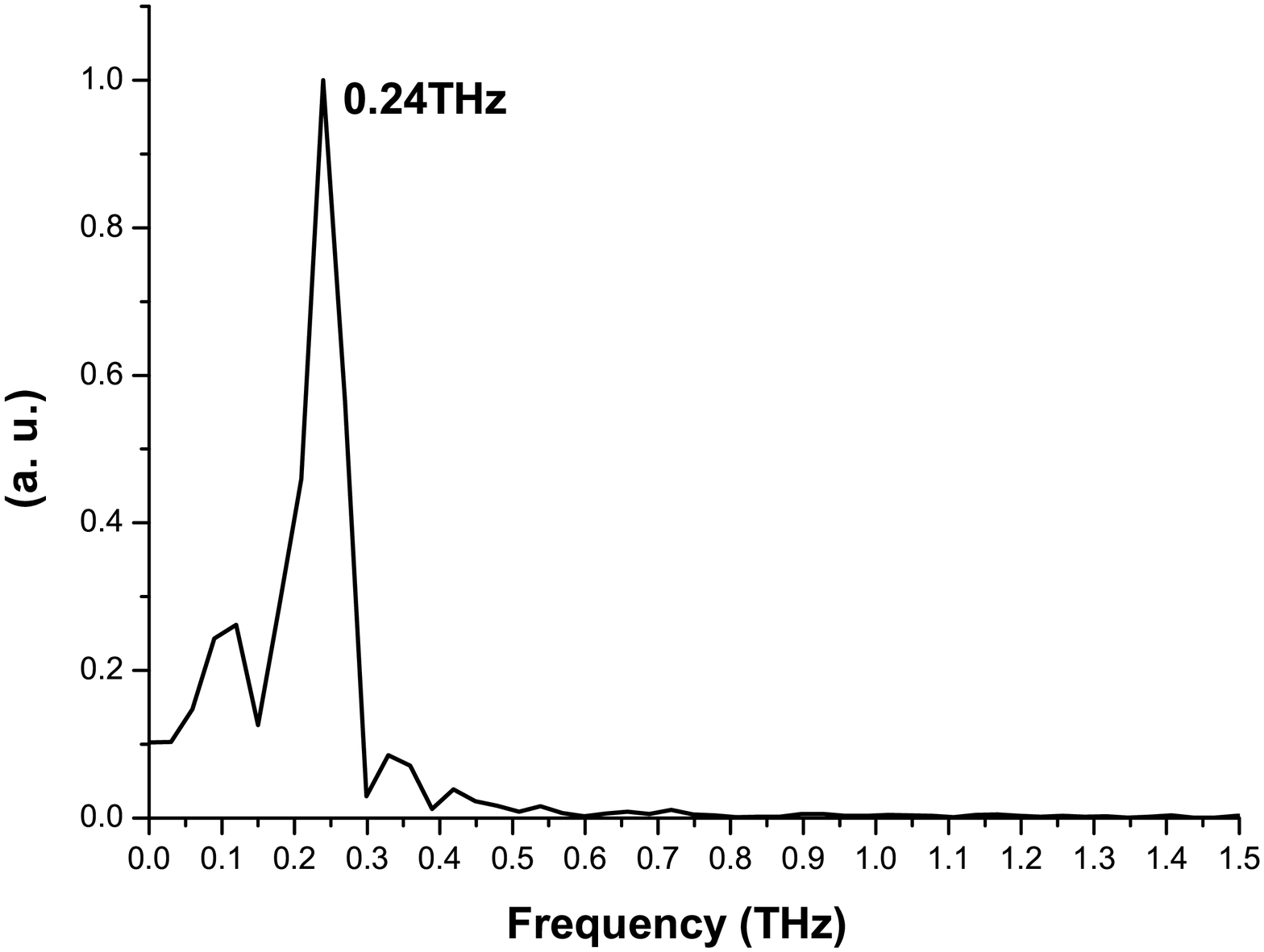}
\figcaption{\label{f6}The power spectrum of the coherent Cerenkov radiation.}
\end{center}
The low frequency hump in the Fig. 6 represents the coherent transition radiation excited by the electron bunch entering the structure. The analytical and simulated results are shown in Table 2.
\end{multicols}
\begin{table}
\begin{center}
\caption{\label{tab:results}Analytical and simulated results of the radiation.}
\begin{tabular}{l c c c c c }
\hline {} & {frequency} & {~$E_{z,max}(0)$~} & {power}  & {pulse length} & {energy}\\
Analytical & 0.241THz  & 49 MV/m & 2.05 MW & 56.49 ps & 0.115 mJ \\
Simulated & 0.240THz  & 50 MV/m & 2.13 MW & 57.04 ps & 0.12 mJ \\
\hline
\end{tabular}
\end{center}
\end{table}
\begin{multicols}{2}
\section{Summary and discussion}
A scheme of compact THz radiation source based on the photocathode rf gun is proposed. The coherent Cherenkov radiation is analytically and numerically studied, which is driven by a subpicosecond electron bunch traveling along the axis of a hollow cylindrical dielectric-lined waveguide. We estimate that 2 MW peak power CCR at 0.24 THz wavelength can be produced using the electron beam capably obtained by the worldwide BNL type photocathode rf gun. When the gun is operated at 120 Hz repetition, above 10 mW average power can be gained, which can be potentially improved by times by extending the length of the DLW.

\end{multicols}

\vspace{-1mm}
\centerline{\rule{80mm}{0.1pt}}
\vspace{2mm}

\begin{multicols}{2}

\end{multicols}

\clearpage

\end{document}